\documentclass[twocolumn,floats,showpacs,amsmath,amssymb,prb]{revtex4}

\usepackage{graphicx}

\begin{document}

\title{Rare-gas solids under pressure: A path-integral Monte Carlo simulation}
\author{Carlos P. Herrero}
\author{Rafael Ram\'{\i}rez}
\affiliation{Instituto de Ciencia de Materiales,
         Consejo Superior de Investigaciones Cient\'{\i}ficas (CSIC), 
         Campus de Cantoblanco, 28049 Madrid, Spain }
\date{\today}

\begin{abstract}
Rare-gas solids (Ne, Ar, Kr, and Xe) under hydrostatic pressure up to 
30 kbar have been studied by  path-integral Monte Carlo simulations in the 
isothermal-isobaric ensemble. 
Results of these simulations have been compared with available experimental
data and with those obtained from a quasiharmonic approximation (QHA). 
This comparison allows us to quantify the overall anharmonicity of the 
lattice vibrations and its influence
on several structural and thermodynamic properties of rare-gas solids.
The vibrational energy increases with pressure, but this increase is
slower than that of the elastic energy, which dominates at high pressures.
In the PIMC simulations, the vibrational kinetic energy is found to be 
larger than the corresponding potential energy, and the relative difference 
between both energies decreases as the applied pressure is raised.
The accuracy of the QHA increases for rising pressure. 
\end{abstract}

\pacs{67.80.-s, 62.50.+p, 65.40.De, 05.10.Ln}

\maketitle

\section{Introduction}
The importance of anharmonic effects in solids has been recognized
long ago, as they are responsible for well-known phenomena such as thermal 
expansion, pressure dependence of the compressibility, phonon couplings, 
as well as isotope dependence of structural properties and melting 
temperature.\cite{as76,sr90}
This kind of effects have been studied both theoretically and experimentally
for rare-gas solids since many years,\cite{po64,kl76} because these are simple 
systems allowing fruitful comparisons between theory and experiment.
The interatomic forces are weak, short range, and rather well understood,
so that critical tests of appropriate theories by their ability to predict
properties of actual rare-gas crystals are relatively simple.
In particular, their thermodynamic properties are interesting due to the 
large anharmonic contributions to their lattice dynamics. 

From a theoretical point of view, anharmonic effects in solids
have been traditionally studied by using approaches such
as the so-called quasiharmonic approximation (QHA).\cite{as76,sr90}
In this approach, frequencies of vibrational modes are assumed to change 
with crystal volume, and for given volume and temperature, the solid 
is supposed to be harmonic.\cite{va98,de02} 
However, the QHA does not deal with phonon interaction effects, which can 
be treated by perturbation theory\cite{wa72} when anharmonicity is not large, 
or by different self-consistent phonon theories for larger 
anharmonicities.\cite{gi68,go69,kl73,pa82}
A different theoretical procedure is the Feynman path integral
method,\cite{kl90} which is well-suited to study thermodynamic 
properties of solids at temperatures lower than the Debye temperature 
$\Theta_D$, where the quantum nature of the atomic nuclei is relevant.
The combination of path integrals with Monte Carlo (MC) sampling
enables us to carry out quantitative and nonperturbative 
studies of anharmonic effects in solids.
 The path-integral Monte Carlo 
(PIMC) technique has been applied earlier to study several properties
of rare-gas solids.\cite{cu93,ne00,ch02,ne02,mu95,he02,he03a} 
In particular, it has predicted
kinetic-energy values in good agreement with experimental data.\cite{ti03,cu97}
An effective-potential Monte Carlo theory\cite{ac95a,ac00} has been
also applied to study thermal and elastic properties of solid neon.

Anharmonic effects increase appreciably with temperature. This is now 
well-known and has been explained quantitatively for rare-gas 
solids.\cite{va98,ac00} In recent years, the effect of pressure on these
solids has attracted much attention from both
experimentalists\cite{gr86,he89,sh01,er02} and theorists.\cite{ne00,ii01,de02,ts02}
The influence of pressure on the anharmonicity of lattice vibrations
is, however, not well understood. 
It has been recently suggested that pressure causes a decrease in this
anharmonicity,\cite{ka03,la04} in line with earlier observations that
the accuracy of the QHA increases as pressure is raised.\cite{po72} 
It has been also argued that at high pressures, thermodynamic properties
of solids can be well described by classical calculations, i.e., dealing
with the atoms as classical oscillators in a given potential.\cite{et74}
This seems to
be at first sight contradictory with the fact that pressure induces a 
larger zero-point vibrational energy of the solid. These questions are
indeed related with the ratio of the vibrational energy to the whole 
internal energy on one side, and with the size of the ``intrinsic'' 
anharmonicity (further than the QHA) of the lattice vibrations, on the 
other side.     

In this paper, we study structural and thermodynamic properties of rare-gas 
solids under pressure. This allows us to study properties of these solids
along well-defined isotherms, and to analyze changes in anharmonic effects
due to the repulsive (for compression) and attractive (for dilation, i.e.,
negative pressure) parts of the interatomic potential.
The interatomic interaction is described by a Lennard-Jones potential.
Results of the PIMC simulations are compared with those yielded by a 
quasiharmonic approximation with the same interatomic potential.
This approach will help us to quantify the influence of the ``intrinsic''
anharmonicity on the considered properties.

\begin{table*}
\caption{Parameters $\sigma$ and $\epsilon$ of the Lennard-Jones potential
employed in this work and average isotopic mass $\langle M \rangle$ for
rare gases.  Calculated zero-temperature properties of rare-gas solids at
zero pressure are also given: lattice parameter $a$, zero-point
vibrational energy $E_{\rm vib}$, and elastic energy $E_{\rm el}$ per atom,
as derived from PIMC simulations.
}
\vspace{3mm}
\begin{tabular}{ccccccccc}
 \,\, Element \,\,& \,\,$\sigma$ (\AA) \,\,& \,\,$\epsilon$ (meV) \,\,
   & $\langle M \rangle$ (amu) & \,\,\, $a$ (\AA)\,\,\,
   & \,\,$E_{\rm vib}$ (meV)\,\,  & \,\, $E_{\rm el}$
(meV)\,\, \\
\colrule
  Ne & 2.782 & 3.084 &  20.18  &   4.4631  &  6.33   &  1.20  \\
  Ar & 3.404 & 10.32 &  39.95  &   5.3115  &  7.99   &  0.43   \\
  Kr & 3.638 & 14.17 &  83.80  &   5.6458  &  6.27   &  0.18   \\
  Xe & 3.961 & 19.91 & 131.30  &   6.1316  &  5.54   &  0.10   \\
\end{tabular}
\end{table*}

The paper is organized as follows.  In Sec.\,II, the
computational method is described. In Sec.\,III, we 
present results for energy, heat capacity, lattice parameter, 
and bulk modulus. Finally, Sec.\,IV includes a discussion of the results 
and the conclusions.

\section{Method}

\subsection{Path-integral Monte Carlo}
Rare-gas atoms were treated as quantum particles interacting
through a Lennard-Jones potential:
$V(r) = 4 \epsilon [ (\sigma / r)^{12} - (\sigma / r)^6 ]$,
with parameters $\epsilon$ and $\sigma$ given in Table I, which were
employed in earlier simulations of this kind of crystals.\cite{he02,he03a}
In this Table we also give the average atomic mass of rare gases used
in the calculations, as well as low-temperature properties of the studied
crystals at zero pressure, as derived from PIMC simulations (see below). 
Lennard-Jones-type potentials have been employed in recent years to model
the atomic interaction in rare-gas solids.\cite{ne02,mu95,ac00,ac95a}
Although more sophisticated interaction potentials have been developed,
they do not seem to be significantly superior to Lennard-Jones potentials 
in accounting for the experimental data.\cite{ne02,ac00} This is not the
case when one considers rare-gas solids under high pressure,
where three-body potentials are necessary (see below). For this reason, our
calculations are restricted to pressures not higher than 30 kbar.

Equilibrium properties of rare-gas solids have been calculated by PIMC 
simulations in the isothermal-isobaric ensemble ($NPT$).
 Simulations have been performed on $5 \times 5 \times 5$ cubic supercells
of the face-centered-cubic unit cell, including 500 rare-gas atoms, and
assuming periodic boundary conditions.
To check the convergence of our results with system size, some MC simulations 
were carried out for other supercell sizes, including $7 \times 7 \times 7$ 
supercells. We found that finite-size effects for $5 \times 5 \times 5$ 
supercells are negligible for the quantities studied here (they are 
smaller than the statistical noise).

In the path-integral formulation of statistical mechanics, the partition 
function is evaluated through a discretization of the density matrix
along cyclic paths, composed of a finite number $N_{\rm Tr}$ (Trotter number) 
of `imaginary-time' steps.\cite{kl90} In the numerical simulations, 
this discretization gives rise to the appearance of $N_{\rm Tr}$ replicas 
for each quantum particle.
In this way, the implementation of this method is based on an isomorphism
between the quantum system and a classical one, obtained by
replacing each quantum particle (here, atomic nucleus)
by a cyclic chain of $N_{\rm Tr}$ classical particles, connected
by harmonic springs with a temperature-dependent constant.
Details on this computational method can be found
elsewhere.\cite{gi88,ce95,no96} 

To have a nearly constant precision for the simulation results
at different temperatures, we considered a Trotter number that
scales as the inverse temperature. At a given $T$, the actual
value $N_{\rm Tr}$ required to obtain convergence of the results depends
on the Debye temperature $\Theta_D$ (higher $\Theta_D$
needs larger $N_{\rm Tr}$). For the simulations at zero pressure, we have taken 
$N_{\rm Tr} T = 250$ K for solid Ar and $N_{\rm Tr} T = 200$ K for the other 
rare-gas solids ($\Theta_D \sim 90$ K for Ar vs $\sim 70$ K for Ne, Kr, 
and Xe).  Since vibrational
frequencies (and the associated Debye temperature) increase as
the applied pressure is raised, the Trotter number has to be correspondingly 
increased. Thus, for a given solid and an applied pressure we have taken
$N_{\rm Tr}$ values roughly proportional to the zero-point vibrational energy at 
the considered pressure. This means that $N_{\rm Tr}$ is increased by a factor of 
about two for Ar, Kr and Xe (about three for Ne) when pressure rises from zero
to 30 kbar. Thus, the computational time required to carry out PIMC
simulations rises (a) as temperature is lowered ($\varpropto 1/T$), and (b) as
pressure is raised [$\varpropto E_{\rm vib}(0)$, zero-point vibrational
energy]. For example, a PIMC simulation for solid Ar at 5 K and zero pressure 
($N_{\rm Tr} = 50$, $N = 500$) is equivalent in computational time to a classical 
MC simulation for $N_{\rm Tr} N$ = 25000 atoms. This number increases by a factor of
two at the same temperature and $P$ = 30 kbar.

Sampling of the configuration space has been carried out by the Metropolis
method at temperatures between 5 K and the triple-point temperature $T_{\rm tp}$ 
of the different solids, as well as at pressures up to 30 kbar.
 For given temperature and pressure, a typical run consisted
of the generation of $2 \times 10^4$ quantum paths per atom for system
equilibration, followed by $3 \times 10^5$ paths per atom
for the calculation of ensemble average properties.
Other technical details are the same as those used in Refs.
\onlinecite{he02,he03a}.

The isothermal bulk modulus $B$ can be obtained in the $NPT$ ensemble from 
the mean-square fluctuations in the lattice parameter, $\sigma_a^2$.  
In this ensemble, fluctuations in the volume $V$ of the simulation cell are 
given by\cite{la80}  $\sigma_V^2 = V k_B T / B$, and therefore
\begin{equation}
      B = \frac{k_B T}{9 L^3 a \sigma_a^2}   \; ,
\label{bulk}
\end{equation}
where $L$ is the side length of the simulation cell in units of the lattice
parameter (here, $L = 5$). 

\subsection{Quasiharmonic approximation}

In the following section, results of PIMC simulations are
compared with those derived from a QHA.  This approximation is based on a 
renormalization of the phonon frequencies with volume, and for a given
volume the solid is assumed to be harmonic.\cite{as76,sr90}
This volume dependence of phonon frequencies is usually described by 
a mode-dependent Gr\"uneisen parameter\cite{as76,sr90}
$\gamma_n({\bf q}) = - \partial \ln \omega_n({\bf q}) /
 \partial \ln V $, where $\omega_n({\bf q})$ are
the frequencies of the $n$th mode in the crystal, and
for small volume changes $\gamma_n({\bf q})$ is assumed to be
constant for each mode.  However, for the QHA calculations
presented here, we have not employed this description based on 
Gr\"uneisen parameters.  Instead of this, we calculate directly 
the actual (harmonic) vibrational frequencies for each crystal
volume, by diagonalizing the corresponding dynamical matrix.

For direct comparison with results of PIMC simulations,
we have employed for the QHA the same supercell as for the simulations, 
i.e., a $5\times5\times5$ supercell with periodic boundary conditions.
This means that only the ${\bf q}=0$ modes in the Brillouin zone
of the supercell are included in the calculation, since modes with 
${\bf q} \neq 0$ violate the periodic boundary conditions.
Then, the total number of vibrational modes in the QHA is 1497, i.e.,
three times the number of rare-gas atoms in the supercell minus
three translational degrees of freedom.
The point group symmetry of the simulation cell imposes that 
many of these normal frequencies are degenerated. 
The number of normal modes that are not symmetry equivalent is 72
for the supercell employed here.
For each temperature, we calculated the free energy as a function
of volume, with the corresponding phonon frequencies.
The lattice parameter was changed in steps of $10^{-3}$ \AA, and from 
the volume derivative of the free energy  we derived the equilibrium volume
as a function of pressure.

\section{Results}

\subsection{Energy}

Once defined an interatomic potential, the internal energy of a solid,
$E(V,T)$, at given volume and temperature can be written as:
\begin{equation}
  E(V,T) = E_0 + E_{\rm el}(V) + E_{\rm vib}(V,T)   \, ,
\label{evt}
\end{equation}
where $E_0$ is the minimum potential energy for the (classical) crystal 
at $T$ = 0, $E_{\rm el}(V)$ is the elastic energy, and $E_{\rm vib}(V,T)$ 
is the vibrational energy. Since we are working in the isothermal-isobaric
ensemble, it is understood that the volume is implicitly given by the
applied pressure, i.e., $V = V(P)$.
For a given volume $V$, the classical energy at $T = 0$ increases by an 
amount $E_{\rm el}(V)$ with respect to the minimum energy $E_0$. 
This elastic energy $E_{\rm el}$ depends only on volume, but at finite
temperatures and for the real (quantum) solids, it depends implicitly
on $T$ due to the temperature dependence of $V$ (thermal expansion).
The elastic energy $E_{\rm el}(V)$ represents a nonnegligible part of the
internal energy, even at zero pressure. For example, in Ne it is found to be 
1.2 and 2.1 meV per atom at 5 and 24 K, respectively.
These values are smaller for the other rare gases, as shown in Table I.

\begin{figure}
\vspace{-4.0cm}
\hspace*{0.4cm}
\includegraphics[width= 10cm]{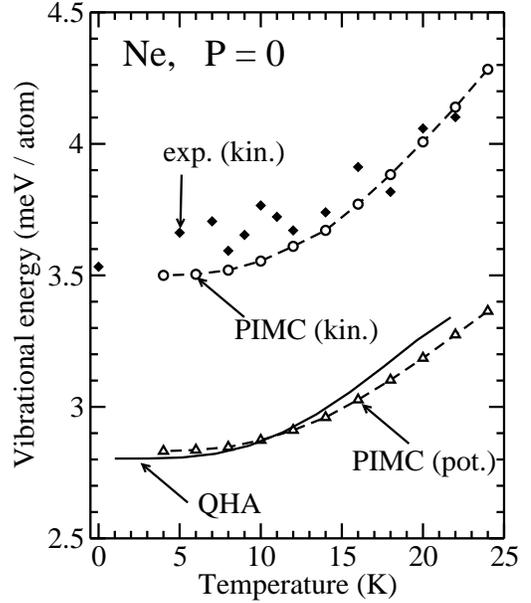}
\vspace{-1.0cm}
\caption{\label{f1}
Temperature dependence of the vibrational energy of solid neon.
Circles and triangles correspond to kinetic and potential energy,
respectively, as derived from PIMC simulations. Error bars of the
simulation results are less than the symbol size. Dashed lines are
guides to the eye.  The solid line is the result of the QHA.
Diamonds are kinetic-energy data for solid Ne, obtained by Timms
{\em et al.}\cite{ti03} from neutron Compton scattering (the
point at $T$ = 0 is an extrapolation given by these authors).
}
\end{figure}

The vibrational energy, $E_{\rm vib}(V,T)$, depends explicitly on both,
$V$ and $T$, and can be obtained by subtracting the elastic
energy from the internal energy.
Values of $E_{\rm vib}$ derived from our PIMC simulations for $T \to 0$ 
and $P$ = 0 are given in Table I.
Path-integral Monte Carlo simulations allow us to obtain
separately the kinetic, $E_k$, and potential energy, $E_p$,
associated to the lattice vibrations.\cite{gi88}
Both energies are shown in Fig. \ref{f1} for solid Ne as a function of
temperature at $P = 0$.  Circles and triangles correspond to the 
vibrational kinetic and potential energy, respectively. Our results 
for the kinetic energy are close to those derived earlier from PIMC 
simulations with Lennard-Jones\cite{cu93,ti96,cu97} and
Aziz\cite{ti96} interatomic potentials.
For comparison, we present also in Fig. \ref{f1} values of the kinetic energy
of Ne atoms, derived by Timms {\em et al.}\cite{ti03}
from neutron Compton scattering in solid neon (black diamonds). 
According to the results of our PIMC simulations, $E_k$
is larger than $E_p$ by about 20\%.
The QHA predicts potential (and kinetic) energy values (solid line) which 
are close to the vibrational potential energy derived from our PIMC 
simulations.
Something similar happens for the other rare-gas solids, with $E_k > E_p$ 
for all temperatures and pressures studied here.
The difference $E_k - E_p$ decreases for increasing atomic mass, and 
at $T$ = 5 K and zero pressure, we find $E_k - E_p$ =
0.67 and 0.059 meV/atom for Ne and Xe, respectively. These energy differences 
increase slowly as pressure rises, and take values of 0.75 and 0.060 meV/atom 
for Ne and Xe at 30 kbar. 

\begin{figure}
\vspace{-4.0cm}
\hspace*{0.4cm}
\includegraphics[width= 10cm]{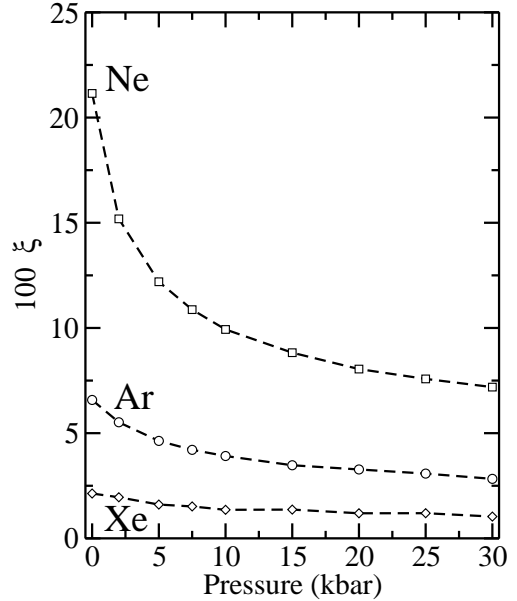}
\vspace{-1.0cm}
\caption{\label{f2}
Anharmonicity parameter $\xi$ for rare-gas crystals as a function of
applied hydrostatic pressure. Symbols indicate results of PIMC
simulations at $T$ = 5 K.  From top to bottom: Ne, Ar, and Xe.
Error bars are less than the symbol size.  Lines are guides to the eye.
Results for Kr lie between those for Ar and Xe, and are not shown for
clarity of the figure.
}
\end{figure}

A quantitative estimation of the overall anharmonicity of the atom
vibrations is given by the parameter\cite{mu95} 
$\xi = 2 (E_k - E_p) / (E_k + E_p)$,
which should be zero for a harmonic solid at any temperature,
as follows from the virial theorem. For rare-gas solids,
it was shown earlier\cite{he03a} that $\xi$ increases as temperature rises, as
expected for larger anharmonicity.
In  Fig. \ref{f2} we show the pressure dependence of the parameter $\xi$
for different rare-gas solids at $T$ = 5 K, as derived from PIMC simulations.
One observes that $\xi$ decreases as pressure is raised. The relative
change in this parameter is largest for Ne, for which it decreases by a
factor of about 3.5. For Xe, it changes by a factor $\approx 2$. 

\begin{figure}
\vspace{-0.5cm}
\hspace*{0.5cm}
\includegraphics[width= 13cm]{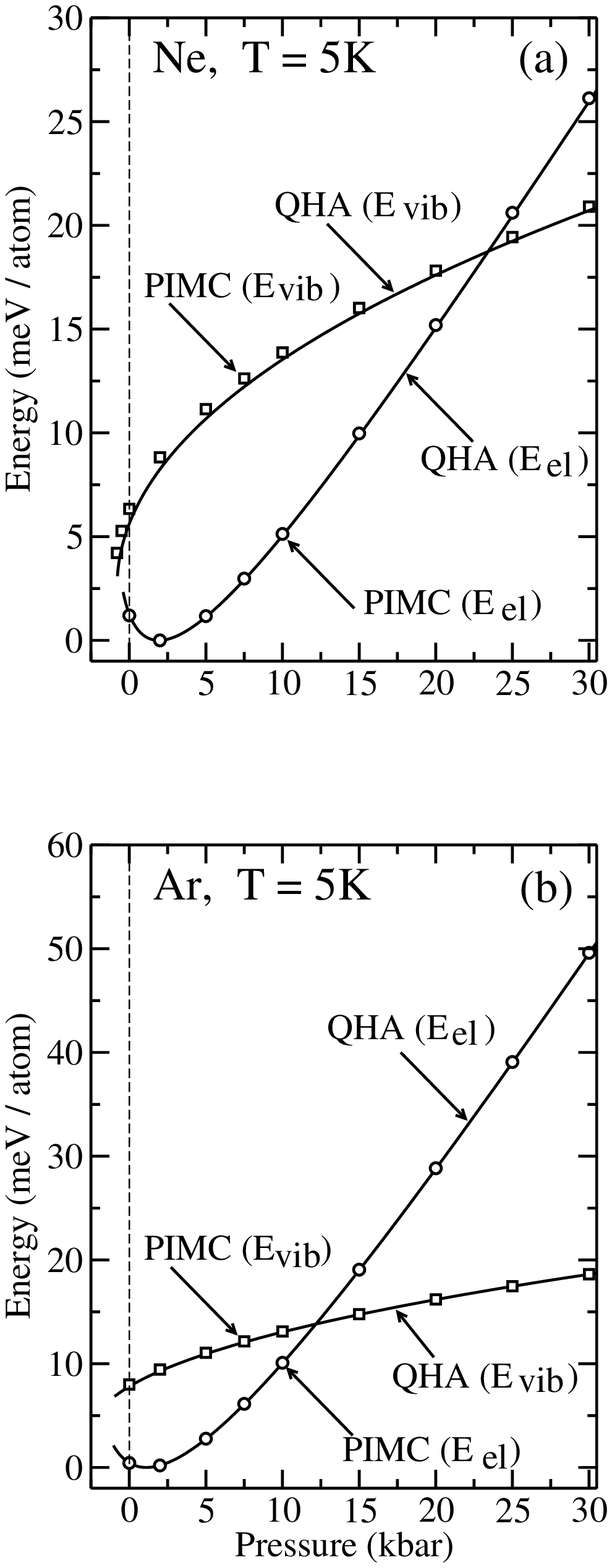}
\vspace{-0.9cm}
\caption{\label{f3}
Vibrational and elastic energy of solid Ne (a) and Ar (b) at $T$ = 5 K,
as a function of pressure. Symbols show results of PIMC simulations:
squares, vibrational energy; circles, elastic energy. Solid lines are
results of the quasiharmonic approximation.
Error bars of the simulation results are less than the symbol size.
}
\end{figure}

The elastic energy $E_{\rm el}$ increases fast as pressure rises.
In Fig. \ref{f3} we display the pressure dependence of $E_{\rm el}$ for (a) Ne
and (b) Ar, obtained from PIMC (open squares) and
QHA (solid lines). For comparison we also
present results for the vibrational energy (kinetic plus potential)
obtained by both procedures.  
As shown above, $E_k$ for Ne at zero pressure is about 20\%
larger than $E_p$, and thus the whole $E_{\rm vib}$
derived from PIMC is 10\% larger than that found in the QHA
($E_p$ is nearly the same for both methods). This relative difference
decreases as pressure increases, and is almost inobservable on the scale of
Fig. \ref{f3}.
Note that close to $P = 0$, $E_{\rm el}$ decreases as $P$ rises, reaches
a minimum and then it increases continuously for larger applied pressures.
This is due to the fact that at $P = 0$ the crystal is expanded with
respect to the volume giving the minimum  potential energy.
With an applied pressure of about 2 kbar, solid neon reaches the
volume corresponding to the classical minimum (giving $E_{\rm el} = 0$), and
for larger pressures the volume is further reduced, with an increase in
$E_{\rm el}$. Something similar happens for Ar and the other rare-gas solids,
although it is less appreciable on the scale of Fig. \ref{f3}(b).
For solid Ne we find that the vibrational energy is larger than $E_{\rm el}$ at
pressures $P < 23$ kbar. However, $E_{\rm el}$ rises faster with
pressure, and becomes the dominant part of the internal energy for larger
pressures. This behavior is qualitatively similar for the other rare-gas
solids, as shown in Fig. \ref{f3}(b) for Ar. The main difference
is that the elastic energy for Ar increases with pressure faster than for
Ne, and becomes larger than $E_{\rm vib}$ for $P > 12$ kbar.
In this context, the main point concerning large pressures is that 
$d E_{\rm el} / d P > d E_{\rm vib} / d P$. This means that the ratio
$E_{\rm el} / E_{\rm vib}$ grows for increasing pressure, and eventually
the vibrational energy becomes a small correction to the internal energy. 

Note that in Fig. \ref{f3} we have included some points at negative 
pressure (that is, solids under tension). This region with $P < 0$ was
studied earlier for rare-gas solids by PIMC simulations.\cite{he03b} 
It was found, in particular, that at $T$ = 5 K solid Ne and Ar 
are metastable until reaching the corresponding spinodal pressures
of --0.9 and --2.5 kbar, where they become mechanically unstable.
Here we only comment that energy results obtained from the QHA follow 
those yielded by PIMC simulations also in this region of negative 
pressures. 

\subsection{Heat capacity}

\begin{figure}
\vspace{-4.0cm}
\hspace*{0.4cm}
\includegraphics[width= 10cm]{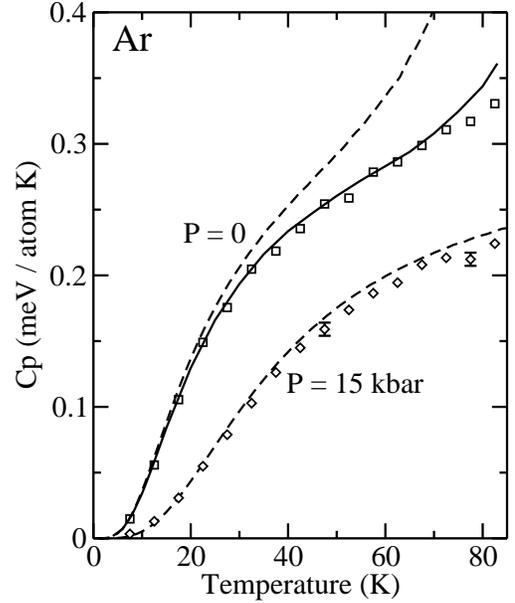}
\vspace{-1.0cm}
\caption{\label{f4}
Heat capacity, $C_p$, of solid argon as a function of temperature at
$P$ = 0 and 15 kbar.  Symbols: PIMC simulations; dashed lines:
quasiharmonic approximation; solid line: experimental data obtained by
Flubacher {\em et al.}\cite{fl61} at atmospheric pressure. Error bars
of the simulation results for $P = 0$ are on the order of the symbol size.
}
\end{figure}

We have calculated the heat capacity $C_p$ of rare-gas solids 
for several pressures as a numerical derivative of the enthalpy, 
$H = E + P V$, with respect to the temperature.
In Fig. \ref{f4} we present results derived from PIMC simulations
(symbols) for solid argon at $P$ = 0 and 15 kbar.  Results of the 
quasiharmonic approximation are shown as dashed lines. For comparison,
we also present experimental results obtained by Flubacher 
{\em et al.}\cite{fl61} for argon at atmospheric pressure.
Results of PIMC simulations at $P = 0$ agree well with experimental
data for both Ne and Ar (data for Ne were given elsewhere\cite{he02} and are 
not shown here), except close to $T_{\rm tp}$, where the 
simulation results are lower than the experimental ones in both cases. 
However, for $T \lesssim T_{\rm tp}$ there can be a systematic 
error in the experimental results ($\sim 10\%$ at $T_{\rm tp}$), 
due to partial vaporizing of the solid.\cite{so71}
Our simulations were carried out for perfect crystals (without vacancies),
and thus we cannot conclude at this point if the observed difference is 
caused by a failure of the Lennard-Jones potential employed here, or by
the experimental uncertainty near $T_{\rm tp}$.

For zero pressure, the QHA gives results close to those of PIMC 
at low temperatures ($T < 5$ K for Ne and $T < 20$ K for Ar), 
but clearly overestimates the heat capacity at higher temperatures. 
The relative error of the QHA (compared with PIMC) increases
as temperature is raised, as expected for an increase in (intrinsic) 
anharmonicity due to thermal effects.   
In the case of Ne, this approximation breaks down at $T$ = 22.6 K, 
due to a mechanical instability (the solid becomes unstable in the QHA at 
a temperature lower than the actual $T_{\rm tp}$). 
For Ar this approach predicts a breakdown of the solid at 86.2 K.  
These results are in line with earlier observations on the validity
temperature range of the QHA for these solids\cite{kl76,va98,ha98} 

For increasing pressure, the error of the QHA decreases. 
This is in part due to the renormalization of vibrational frequencies, 
which increase under an applied pressure, and therefore shift the increase in
$C_p$ to higher temperatures. Moreover, the relative contribution 
of the vibrational energy to the internal energy is reduced as pressure
rises, and in consequence the intrinsic anharmonicity of the
vibrational modes (not captured by the QHA) becomes less relevant for
the heat capacity.

\subsection{Lattice parameter}

At $T = 0$, the difference $\Delta a(0) = a(0) - a_{\rm cl}(0)$
between the actual lattice parameter $a(0)$ and that corresponding to the
minimum potential energy of the (classical) crystal, $a_{\rm cl}(0)$,
decreases as the atomic mass rises and quantum effects become less 
relevant.\cite{mu95,he01}
From our PIMC simulations we found at $P = 0$
that $\Delta a(0)$ ranges from 0.174 \AA\ for Ne to 0.025 \AA\ for Xe.
Calculated values for $a(0)$ at zero pressure are given in Table I.
The temperature dependence of the lattice parameter derived from this kind of 
PIMC simulations agrees with experimental data for rare-gas 
solids.\cite{he02,he03a}  Such an agreement is also found at low
temperatures in the pressure range considered here ($P \le 30$ kbar).
In this pressure range, the QHA predicts lattice parameters which follow 
closely those given by the simulations (see below the discussion on the bulk 
modulus).

For large pressures, it is known that the description of rare-gas
solids with effective interatomic potentials, requires the consideration
of three-body terms\cite{gr86,lo97,ne00} to reproduce the actual equation of 
state $P$--$V$. We have checked that the Lennard-Jones potential considered
here predicts $P$--$V$ isotherms for solid Ar close to the experimental 
ones\cite{sh01} up to pressures on the order of 50 kbar. 
For larger pressures, this pair potential yields volumes larger than
the real ones. 

\begin{figure}
\vspace{-4.0cm}
\hspace*{0.4cm}
\includegraphics[width= 10cm]{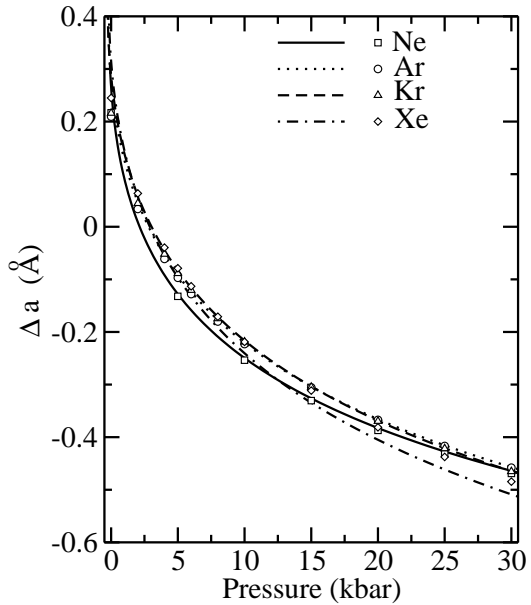}
\vspace{-1.0cm}
\caption{\label{f5}
Pressure dependence of the increment in lattice parameter,
$\Delta a = a - a_{\rm cl}(0)$, with respect to the ideal (classical)
crystal with minimum potential energy. Symbols and lines indicate results
of PIMC simulations and QHA along a given isotherm for each solid:
Ne at 20 K (squares, solid line), Ar at 80 K (circles, dotted line),
Kr at 110 K (triangles, dashed line), and Xe at 160 K (diamonds,
dashed-dotted line).
}
\end{figure}

Even though thermal effects on the crystal volume change in magnitude for 
different rare-gas solids at low temperatures ($T \lesssim 5$ K), these 
differences are less important at higher temperatures. If one takes as a
reference the classical lattice parameter $a_{\rm cl}(0)$, 
the difference $\Delta a = a - a_{\rm cl}(0)$ at temperatures close to 
$T_{\rm tp}$ of each solid amounts to $\approx 0.25$ \AA, as derived from 
our PIMC simulations.
This difference $\Delta a$ is plotted in Fig. \ref{f5} for the different solids
as a function of pressure.  For each solid we present PIMC results 
along an isotherm: Ne, $T = 20$ K; Ar, 80 K; Kr, 110 K;
Xe, 160 K. These results for the different rare-gas solids follow roughly 
the same pressure dependence. Note that at $P = 0$, $\Delta a$ is positive, as a
consequence of zero-point and thermal lattice expansion. Under applied
pressure $\Delta a$ decreases, reaching $\Delta a = 0$ for a pressure
$P \approx 2$ kbar, and is negative for larger $P$.  
Lines in this figure show results of the QHA, which follow closely those of
PIMC. The main difference between both sets of results appears 
for Xe at pressures higher than 10 kbar (diamonds and dashed-dotted lines).
We conclude that the difference $\Delta a$ at temperatures near $T_{\rm tp}$
follows a pressure dependence similar for all rare-gas solids considered
here.

\subsection{Bulk modulus}

The isothermal bulk modulus $B$ of rare-gas solids has been calculated 
from our PIMC simulations by using Eq.\,(\ref{bulk}).
The temperature dependence of $B$ at zero pressure was studied 
earlier\cite{he02,he03a} and will not be presented here. 
We only note that results obtained from this kind of simulations for Ar, Kr,
and Xe showed good agreement with experimental data up to temperatures
close to $T_{\rm tp}$.\cite{he02,he03a} 
Results found for the bulk modulus of neon are lower than the experimental 
data at $T < 18$ K,  and at higher $T$ the experimental $B$
decreases faster than that derived from the simulations.
This seems to be a general problem of this kind of calculations with 
Lennard-Jones and Tang-Toennis-type interatomic potentials.\cite{ac00,ta84}

\begin{figure}
\vspace{-4.0cm}
\hspace*{0.4cm}
\includegraphics[width= 10cm]{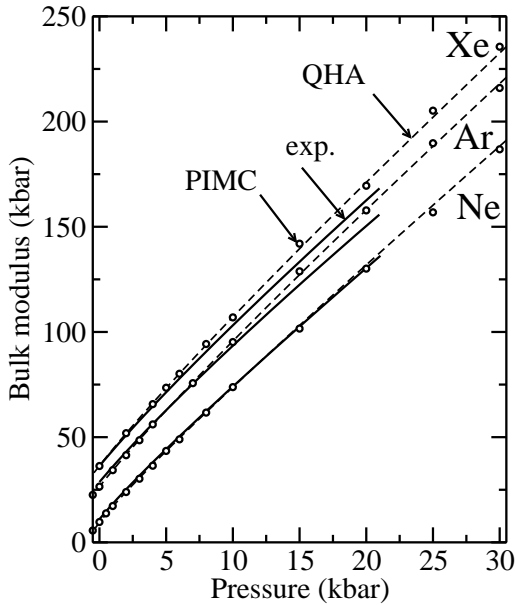}
\vspace{-1.0cm}
\caption{\label{f6}
Isothermal bulk modulus as a function of pressure for rare-gas solids.
Symbols: results of PIMC simulations at $T$ = 5 K, with error bars on
the order of the symbol size. Dashed lines: quasiharmonic approximation
at 5 K.  Solid lines: experimental data by Anderson {\em et al.}\cite{an73}
for Ne, and Anderson and Swenson\cite{an75} for Ar and Xe, at $T$ = 4.2 K.
}
\end{figure}

We now turn into the pressure dependence of the bulk modulus for
these solids.  Results of our PIMC simulations for Ne, Ar, and Xe at 
$T$ = 5 K are plotted in Fig. \ref{f6} (symbols). 
The bulk modulus of Kr (not shown for clarity) lies between those of Ar and Xe.
Dashed lines are results of the QHA, and solid lines
represent experimental data obtained by Anderson {\em et al.}\cite{an73} 
for Ne, and Anderson and Swenson\cite{an75} for Ar and Xe.
The QHA predicts bulk moduli in good agreement with those derived from
PIMC simulations, indicating that this approximation is rather accurate
for predicting the $P$--$V$ equation of state, as well as the derivative 
$\partial P / \partial V$, which gives the bulk modulus.

Our calculated results for Ne agree well with experimental data in the 
pressure region under consideration. For the other rare-gas solids
(including Kr, not shown in Fig. \ref{f6}) our results are larger
than the experimental data for $P \gtrsim$ 10 kbar.
At 20 kbar, our method overestimates the bulk modulus of Ar and Xe by about
5\%. This is a consequence of the interatomic pair potential employed in
our calculations. As indicated above, the limitations of pair potentials for
describing this kind of solids show up as pressure rises, and 
eventually three-body terms are necessary at high pressures.\cite{lo97,ne00} 

\section{Discussion}

Path-integral Monte Carlo simulations give in principle the ``exact'' 
solution to the considered quantum problem, with an accuracy depending on the 
considered Trotter number and the statistical error associated to the
MC sampling. Thus, the Lennard-Jones potential employed here gives a good
description of structural and thermodynamic properties of rare-gas solids 
at zero pressure between $T = 0$  and the triple-point temperature.
The equation of state $P-V$ is well described by this potential in the pressure
range considered here ($P < 30$ kbar). The bulk modulus of Ne is well
reproduced, and for heavier rare gases PIMC simulations give $B$ values
larger than experiment at $P \gtrsim 10$ kbar.

In all solids considered here, and for different pressures and
temperatures, we have found $E_k > E_p$. The overall anharmonicity
has been measured by the parameter $\xi$, which for given $T$ and
$P$ decreases as the atomic mass is raised. 
Although this parameter is by no means a unique and absolute measure of the 
whole anharmonicity, for a family of similar materials it is useful to give
us a quantitative estimation of anharmonic 
effects as a function of atomic mass, temperature, and pressure.
In the limit $P = 0$
and $T \to 0$, $\xi$ changes from 0.21 for Ne to 0.02 for Xe.
This indicates a large anharmonicity of the lattice vibrations in
Ne, even at $T = 0$.
For comparison, we note that covalent solids at low temperatures
show much lower values of $\xi$. Thus, for diamond, silicon, and germanium
one finds for $T / \Theta_D \ll 1$, differences between $E_p$ and $E_k$
smaller than $1\%$, and at $T \sim \Theta_D$ they are less than
$3\%$.\cite{no97,he00a,he01}
Another point of interest is that for such covalent materials the
vibrational $E_p$ was found to be larger than $E_k$, just the opposite to
the trend found for rare-gas solids.
Then, the fact that $E_p < E_k$, obtained for the rare-gas solids studied
here, is not general in solids, and can be due to the particular nature
of the interatomic interactions present in rare-gas (Lennard-Jones) solids.
                                                                                    
  A qualitative understanding of the sizeable increase in kinetic energy
of the rare-gas atoms with respect to the value expected in a QHA can be
obtained by analyzing the changes of kinetic and potential energy by
standard time-independent perturbation methods.  With this purpose,
we consider a one-coordinate perturbed harmonic oscillator with Hamiltonian
$H = p^2 / 2 m + \frac12 m \omega^2 x^2 + A x^3 + B x^4$. The $x^3$ term
does not introduce corrections to the zero-point energy in first
order,\cite{la80} and  the second-order correction is only due to a change
in kinetic energy. The $x^4$ term gives a first-order correction,\cite{la80}
that again is only caused by a kinetic-energy change.
Thus, the leading corrections to the zero-point energy due to both perturbing
terms originate from changes in the kinetic energy.
This is in line with the results presented in Fig. \ref{f1}, which show that
the potential energy derived from PIMC is close to that yielded by
the QHA, and the kinetic energy is far from the QHA result.
This one-coordinate approach can give us only a very qualitative
interpretation of the changes in kinetic and potential energy of the
considered crystals with respect to a quasiharmonic approximation. In fact,
the whole problem is a many-particle one, in which the lattice vibrations
cannot be considered as non-interacting entities when anharmonicities are
present. The coupling between vibrational modes obtained in a harmonic
or quasiharmonic approximation is expected to increase as the crystal
density increases (pressure rises), thus making such approximations
less reliable in the description of the vibrational problem.

For what concerns structural and thermodynamic properties of rare-gas
solids, the QHA becomes more precise as pressure increases. 
This is related to the pressure dependence of the vibrational and elastic 
energies, and is due to two reasons.
First, for high pressures the vibrational energy becomes a small part
of the internal energy, and hence the influence of lattice vibrations 
(and of their anharmonicity) on thermodynamic properties will comparatively 
decrease.  The same happens for the free energy $F$ at finite temperatures, since
its vibrational part $F_{\rm vib}$ becomes small as compared with the whole $F$.
Thus, for $F_{\rm vib} \ll F - E_0$, a QHA describes well the pressure
dependence of the crystal volume, since it is basically given by the elastic 
energy of the solid.
Second, the intrinsic anharmonicity of the lattice vibrations
(as measured by the anharmonicity parameter $\xi$) decreases as
pressure rises (see Fig. \ref{f2}), and eventually becomes negligible for
large $P$ ($\xi \to 0$).
Both arguments go in the same direction of making more accurate the QHA.

Nevertheless, the improved accuracy of the QHA as pressure rises is not a 
particular merit of this approach, 
since the internal energy becomes dominated by the elastic energy 
and the actual description of the vibrational modes 
is not very relevant for thermodynamic properties. It has been 
also argued\cite{et74} that structural properties of solids under 
pressure can be described rather accurately by a classical model for the 
lattice vibrations.
The origin of this is similar to that described above for the decreasing
effect of anharmonicity as $P$ rises, since in this respect 
the actual description of the lattice vibrations by a classical
or a quantum model becomes unimportant for solids under large pressures.
This is not necessarily the case for spectroscopic properties of the
solids under consideration, because vibrational frequencies predicted by a 
QHA or by a classical model are not guaranteed to describe correctly the actual 
ones for high $P$.

In summary, we have analyzed the influence of a hydrostatic pressure on
anharmonic effects of rare-gas solids.
Our results indicate that the validity of the QHA to describe structural
and thermodynamic properties of these solids increases as pressure is
raised. This is mainly a consequence of the relative importance of elastic
and vibrational energy, since the latter becomes increasingly irrelevant
as pressure rises. Therefore, the precision requirements for
a description of the (anharmonic) vibrational modes is reduced with
increasing pressure. In the limit of large pressures, even a classical
description of these modes can be sufficiently precise to predict 
several properties of these solids at low temperatures. 
This is, of course, not the case of vibrational properties, which may 
require the full quantum treatment with the consideration of zero-point
anharmonic effects at high pressures.

We finally note that the extension of the method employed here to study 
solids under larger
pressures, such as those presently reached in experimental studies ($P \sim$
100 GPa), is hampered by the requirement of enlarging enormously the 
Trotter number (and consequently the CPU time) in
PIMC simulations, which has to be increased as the vibrational energy
rises. This problem is particularly important to study vibrational properties,
since for very high pressures, thermodynamic properties can be well described
by neglecting the quantum nature of the atomic nuclei, as usually done
in electronic-structure calculations.

\begin{acknowledgments}
The authors benefitted from discussions with L. M. Ses\'e.
This work was supported by CICYT (Spain) through Grant
No. BFM2003-03372-C03-03.   \\
\end{acknowledgments}

\bibliographystyle{apsrev}

\begin{thebibliography}{49}
\expandafter\ifx\csname natexlab\endcsname\relax\def\natexlab#1{#1}\fi
\expandafter\ifx\csname bibnamefont\endcsname\relax
  \def\bibnamefont#1{#1}\fi
\expandafter\ifx\csname bibfnamefont\endcsname\relax
  \def\bibfnamefont#1{#1}\fi
\expandafter\ifx\csname citenamefont\endcsname\relax
  \def\citenamefont#1{#1}\fi
\expandafter\ifx\csname url\endcsname\relax
  \def\url#1{\texttt{#1}}\fi
\expandafter\ifx\csname urlprefix\endcsname\relax\def\urlprefix{URL }\fi
\providecommand{\bibinfo}[2]{#2}
\providecommand{\eprint}[2][]{\url{#2}}

\bibitem[{\citenamefont{Ashcroft and Mermin}(1976)}]{as76}
\bibinfo{author}{\bibfnamefont{N.~W.} \bibnamefont{Ashcroft}} \bibnamefont{and}
  \bibinfo{author}{\bibfnamefont{N.~D.} \bibnamefont{Mermin}},
  \emph{\bibinfo{title}{Solid State Physics}} (\bibinfo{publisher}{Saunders
  College}, \bibinfo{address}{Philadelphia}, \bibinfo{year}{1976}).

\bibitem[{\citenamefont{Srivastava}(1990)}]{sr90}
\bibinfo{author}{\bibfnamefont{G.~P.} \bibnamefont{Srivastava}},
  \emph{\bibinfo{title}{The Physics of Phonons}} (\bibinfo{publisher}{Adam
  Hilger}, \bibinfo{address}{Bristol}, \bibinfo{year}{1990}).

\bibitem[{\citenamefont{Pollack}(1964)}]{po64}
\bibinfo{author}{\bibfnamefont{G.~L.} \bibnamefont{Pollack}},
  \bibinfo{journal}{Rev. Mod. Phys.} \textbf{\bibinfo{volume}{36}},
  \bibinfo{pages}{748} (\bibinfo{year}{1964}).

\bibitem[{\citenamefont{Klein and Venables}(1976)}]{kl76}
\bibinfo{editor}{\bibfnamefont{M.~L.} \bibnamefont{Klein}} \bibnamefont{and}
  \bibinfo{editor}{\bibfnamefont{J.~A.} \bibnamefont{Venables}}, eds.,
  \emph{\bibinfo{title}{Rare Gas Solids}} (\bibinfo{publisher}{Academic Press},
  \bibinfo{address}{New York}, \bibinfo{year}{1976}).

\bibitem[{\citenamefont{Valle and Venuti}(1998)}]{va98}
\bibinfo{author}{\bibfnamefont{R.~G.~D.} \bibnamefont{Valle}} \bibnamefont{and}
  \bibinfo{author}{\bibfnamefont{E.}~\bibnamefont{Venuti}},
  \bibinfo{journal}{Phys. Rev. B} \textbf{\bibinfo{volume}{58}},
  \bibinfo{pages}{206} (\bibinfo{year}{1998}).

\bibitem[{\citenamefont{Dewhurst et~al.}(2002)\citenamefont{Dewhurst, Ahuja,
  Li, and Johansson}}]{de02}
\bibinfo{author}{\bibfnamefont{J.~K.} \bibnamefont{Dewhurst}},
  \bibinfo{author}{\bibfnamefont{R.}~\bibnamefont{Ahuja}},
  \bibinfo{author}{\bibfnamefont{S.}~\bibnamefont{Li}}, \bibnamefont{and}
  \bibinfo{author}{\bibfnamefont{B.}~\bibnamefont{Johansson}},
  \bibinfo{journal}{Phys. Rev. Lett.} \textbf{\bibinfo{volume}{88}},
  \bibinfo{pages}{075504} (\bibinfo{year}{2002}).

\bibitem[{\citenamefont{Wallace}(1972)}]{wa72}
\bibinfo{author}{\bibfnamefont{D.~C.} \bibnamefont{Wallace}},
  \emph{\bibinfo{title}{Thermodynamics of crystals}} (\bibinfo{publisher}{John
  Wiley}, \bibinfo{address}{New York}, \bibinfo{year}{1972}).

\bibitem[{\citenamefont{Gillis et~al.}(1968)\citenamefont{Gillis, Werthamer,
  and Koehler}}]{gi68}
\bibinfo{author}{\bibfnamefont{N.~S.} \bibnamefont{Gillis}},
  \bibinfo{author}{\bibfnamefont{N.~R.} \bibnamefont{Werthamer}},
  \bibnamefont{and} \bibinfo{author}{\bibfnamefont{T.~R.}
  \bibnamefont{Koehler}}, \bibinfo{journal}{Phys. Rev.}
  \textbf{\bibinfo{volume}{165}}, \bibinfo{pages}{951} (\bibinfo{year}{1968}).

\bibitem[{\citenamefont{Goldman et~al.}(1969)\citenamefont{Goldman, Horton, and
  Klein}}]{go69}
\bibinfo{author}{\bibfnamefont{V.~V.} \bibnamefont{Goldman}},
  \bibinfo{author}{\bibfnamefont{G.~K.} \bibnamefont{Horton}},
  \bibnamefont{and} \bibinfo{author}{\bibfnamefont{M.~L.} \bibnamefont{Klein}},
  \bibinfo{journal}{J. Low Temp. Phys.} \textbf{\bibinfo{volume}{1}},
  \bibinfo{pages}{391} (\bibinfo{year}{1969}).

\bibitem[{\citenamefont{Klein et~al.}(1973)\citenamefont{Klein, Koehler, and
  Gray}}]{kl73}
\bibinfo{author}{\bibfnamefont{M.~L.} \bibnamefont{Klein}},
  \bibinfo{author}{\bibfnamefont{T.~R.} \bibnamefont{Koehler}},
  \bibnamefont{and} \bibinfo{author}{\bibfnamefont{R.~L.} \bibnamefont{Gray}},
  \bibinfo{journal}{Phys. Rev. B} \textbf{\bibinfo{volume}{7}},
  \bibinfo{pages}{1571} (\bibinfo{year}{1973}).

\bibitem[{\citenamefont{Paskin et~al.}(1982)\citenamefont{Paskin, de~Kreiner,
  Shukla, Welch, and Dienes}}]{pa82}
\bibinfo{author}{\bibfnamefont{A.}~\bibnamefont{Paskin}},
  \bibinfo{author}{\bibfnamefont{A.~M.~L.} \bibnamefont{de~Kreiner}},
  \bibinfo{author}{\bibfnamefont{K.}~\bibnamefont{Shukla}},
  \bibinfo{author}{\bibfnamefont{D.~O.} \bibnamefont{Welch}}, \bibnamefont{and}
  \bibinfo{author}{\bibfnamefont{G.~J.} \bibnamefont{Dienes}},
  \bibinfo{journal}{Phys. Rev. B} \textbf{\bibinfo{volume}{25}},
  \bibinfo{pages}{1297} (\bibinfo{year}{1982}).

\bibitem[{\citenamefont{Kleinert}(1990)}]{kl90}
\bibinfo{author}{\bibfnamefont{H.}~\bibnamefont{Kleinert}},
  \emph{\bibinfo{title}{Path Integrals in Quantum Mechanics, Statistics and
  Polymer Physics}} (\bibinfo{publisher}{World Scientific},
  \bibinfo{address}{Singapore}, \bibinfo{year}{1990}).

\bibitem[{\citenamefont{Cuccoli et~al.}(1993)\citenamefont{Cuccoli, Macchi,
  Tognetti, and Vaia}}]{cu93}
\bibinfo{author}{\bibfnamefont{A.}~\bibnamefont{Cuccoli}},
  \bibinfo{author}{\bibfnamefont{A.}~\bibnamefont{Macchi}},
  \bibinfo{author}{\bibfnamefont{V.}~\bibnamefont{Tognetti}}, \bibnamefont{and}
  \bibinfo{author}{\bibfnamefont{R.}~\bibnamefont{Vaia}},
  \bibinfo{journal}{Phys. Rev. B} \textbf{\bibinfo{volume}{47}},
  \bibinfo{pages}{14923} (\bibinfo{year}{1993}).

\bibitem[{\citenamefont{Neumann and Zoppi}(2000)}]{ne00}
\bibinfo{author}{\bibfnamefont{M.}~\bibnamefont{Neumann}} \bibnamefont{and}
  \bibinfo{author}{\bibfnamefont{M.}~\bibnamefont{Zoppi}},
  \bibinfo{journal}{Phys. Rev. B} \textbf{\bibinfo{volume}{62}},
  \bibinfo{pages}{41} (\bibinfo{year}{2000}).

\bibitem[{\citenamefont{Chakravarty}(2002)}]{ch02}
\bibinfo{author}{\bibfnamefont{C.}~\bibnamefont{Chakravarty}},
  \bibinfo{journal}{J. Chem. Phys.} \textbf{\bibinfo{volume}{116}},
  \bibinfo{pages}{8938} (\bibinfo{year}{2002}).

\bibitem[{\citenamefont{Neumann and Zoppi}(2002)}]{ne02}
\bibinfo{author}{\bibfnamefont{M.}~\bibnamefont{Neumann}} \bibnamefont{and}
  \bibinfo{author}{\bibfnamefont{M.}~\bibnamefont{Zoppi}},
  \bibinfo{journal}{Phys. Rev. E} \textbf{\bibinfo{volume}{65}},
  \bibinfo{pages}{031203} (\bibinfo{year}{2002}).

\bibitem[{\citenamefont{M\"user et~al.}(1995)\citenamefont{M\"user, Nielaba,
  and Binder}}]{mu95}
\bibinfo{author}{\bibfnamefont{M.~H.} \bibnamefont{M\"user}},
  \bibinfo{author}{\bibfnamefont{P.}~\bibnamefont{Nielaba}}, \bibnamefont{and}
  \bibinfo{author}{\bibfnamefont{K.}~\bibnamefont{Binder}},
  \bibinfo{journal}{Phys. Rev. B} \textbf{\bibinfo{volume}{51}},
  \bibinfo{pages}{2723} (\bibinfo{year}{1995}).

\bibitem[{\citenamefont{Herrero}(2002)}]{he02}
\bibinfo{author}{\bibfnamefont{C.~P.} \bibnamefont{Herrero}},
  \bibinfo{journal}{Phys. Rev. B} \textbf{\bibinfo{volume}{65}},
  \bibinfo{pages}{014112} (\bibinfo{year}{2002}).

\bibitem[{\citenamefont{Herrero}(2003{\natexlab{a}})}]{he03a}
\bibinfo{author}{\bibfnamefont{C.~P.} \bibnamefont{Herrero}},
  \bibinfo{journal}{J. Phys.: Condens. Matter} \textbf{\bibinfo{volume}{15}},
  \bibinfo{pages}{475} (\bibinfo{year}{2003}{\natexlab{a}}).

\bibitem[{\citenamefont{Timms et~al.}(2003)\citenamefont{Timms, Simmons, and
  Mayers}}]{ti03}
\bibinfo{author}{\bibfnamefont{D.~N.} \bibnamefont{Timms}},
  \bibinfo{author}{\bibfnamefont{R.~O.} \bibnamefont{Simmons}},
  \bibnamefont{and} \bibinfo{author}{\bibfnamefont{J.}~\bibnamefont{Mayers}},
  \bibinfo{journal}{Phys. Rev. B} \textbf{\bibinfo{volume}{67}},
  \bibinfo{pages}{172301} (\bibinfo{year}{2003}).

\bibitem[{\citenamefont{Cuccoli et~al.}(1997)\citenamefont{Cuccoli, Macchi,
  Pedrolli, Tognetti, and Vaia}}]{cu97}
\bibinfo{author}{\bibfnamefont{A.}~\bibnamefont{Cuccoli}},
  \bibinfo{author}{\bibfnamefont{A.}~\bibnamefont{Macchi}},
  \bibinfo{author}{\bibfnamefont{G.}~\bibnamefont{Pedrolli}},
  \bibinfo{author}{\bibfnamefont{V.}~\bibnamefont{Tognetti}}, \bibnamefont{and}
  \bibinfo{author}{\bibfnamefont{R.}~\bibnamefont{Vaia}},
  \bibinfo{journal}{Phys. Rev. B} \textbf{\bibinfo{volume}{56}},
  \bibinfo{pages}{51} (\bibinfo{year}{1997}).

\bibitem[{\citenamefont{Acocella et~al.}(2000)\citenamefont{Acocella, Horton,
  and Cowley}}]{ac00}
\bibinfo{author}{\bibfnamefont{D.}~\bibnamefont{Acocella}},
  \bibinfo{author}{\bibfnamefont{G.~K.} \bibnamefont{Horton}},
  \bibnamefont{and} \bibinfo{author}{\bibfnamefont{E.~R.}
  \bibnamefont{Cowley}}, \bibinfo{journal}{Phys. Rev. B}
  \textbf{\bibinfo{volume}{61}}, \bibinfo{pages}{8753} (\bibinfo{year}{2000}).

\bibitem[{\citenamefont{Acocella et~al.}(1995)\citenamefont{Acocella, Horton,
  and Cowley}}]{ac95a}
\bibinfo{author}{\bibfnamefont{D.}~\bibnamefont{Acocella}},
  \bibinfo{author}{\bibfnamefont{G.~K.} \bibnamefont{Horton}},
  \bibnamefont{and} \bibinfo{author}{\bibfnamefont{E.~R.}
  \bibnamefont{Cowley}}, \bibinfo{journal}{Phys. Rev. Lett.}
  \textbf{\bibinfo{volume}{74}}, \bibinfo{pages}{4887} (\bibinfo{year}{1995}).

\bibitem[{\citenamefont{Grimsditch et~al.}(1986)\citenamefont{Grimsditch,
  Loubeyre, and Polian}}]{gr86}
\bibinfo{author}{\bibfnamefont{M.}~\bibnamefont{Grimsditch}},
  \bibinfo{author}{\bibfnamefont{P.}~\bibnamefont{Loubeyre}}, \bibnamefont{and}
  \bibinfo{author}{\bibfnamefont{A.}~\bibnamefont{Polian}},
  \bibinfo{journal}{Phys. Rev. B} \textbf{\bibinfo{volume}{33}},
  \bibinfo{pages}{7192} (\bibinfo{year}{1986}).

\bibitem[{\citenamefont{Hemley et~al.}(1989)\citenamefont{Hemley, Zha,
  Jephcoat, Mao, Finger, and Cox}}]{he89}
\bibinfo{author}{\bibfnamefont{R.~J.} \bibnamefont{Hemley}},
  \bibinfo{author}{\bibfnamefont{C.~S.} \bibnamefont{Zha}},
  \bibinfo{author}{\bibfnamefont{A.~P.} \bibnamefont{Jephcoat}},
  \bibinfo{author}{\bibfnamefont{H.~K.} \bibnamefont{Mao}},
  \bibinfo{author}{\bibfnamefont{L.~W.} \bibnamefont{Finger}},
  \bibnamefont{and} \bibinfo{author}{\bibfnamefont{D.~E.} \bibnamefont{Cox}},
  \bibinfo{journal}{Phys. Rev. B} \textbf{\bibinfo{volume}{39}},
  \bibinfo{pages}{11820} (\bibinfo{year}{1989}).

\bibitem[{\citenamefont{Shimizu et~al.}(2001)\citenamefont{Shimizu, Tashiro,
  Kume, and Sasaki}}]{sh01}
\bibinfo{author}{\bibfnamefont{H.}~\bibnamefont{Shimizu}},
  \bibinfo{author}{\bibfnamefont{H.}~\bibnamefont{Tashiro}},
  \bibinfo{author}{\bibfnamefont{T.}~\bibnamefont{Kume}}, \bibnamefont{and}
  \bibinfo{author}{\bibfnamefont{S.}~\bibnamefont{Sasaki}},
  \bibinfo{journal}{Phys. Rev. Lett.} \textbf{\bibinfo{volume}{86}},
  \bibinfo{pages}{4568} (\bibinfo{year}{2001}).

\bibitem[{\citenamefont{Errandonea et~al.}(2002)\citenamefont{Errandonea,
  Schwager, Boehler, and Ross}}]{er02}
\bibinfo{author}{\bibfnamefont{D.}~\bibnamefont{Errandonea}},
  \bibinfo{author}{\bibfnamefont{B.}~\bibnamefont{Schwager}},
  \bibinfo{author}{\bibfnamefont{R.}~\bibnamefont{Boehler}}, \bibnamefont{and}
  \bibinfo{author}{\bibfnamefont{M.}~\bibnamefont{Ross}},
  \bibinfo{journal}{Phys. Rev. B} \textbf{\bibinfo{volume}{65}},
  \bibinfo{pages}{214110} (\bibinfo{year}{2002}).

\bibitem[{\citenamefont{Iitaka and Ebisuzaki}(2001)}]{ii01}
\bibinfo{author}{\bibfnamefont{T.}~\bibnamefont{Iitaka}} \bibnamefont{and}
  \bibinfo{author}{\bibfnamefont{T.}~\bibnamefont{Ebisuzaki}},
  \bibinfo{journal}{Phys. Rev. B} \textbf{\bibinfo{volume}{65}},
  \bibinfo{pages}{012103} (\bibinfo{year}{2001}).

\bibitem[{\citenamefont{Tsuchiya and Kawamura}(2002)}]{ts02}
\bibinfo{author}{\bibfnamefont{T.}~\bibnamefont{Tsuchiya}} \bibnamefont{and}
  \bibinfo{author}{\bibfnamefont{K.}~\bibnamefont{Kawamura}},
  \bibinfo{journal}{J. Chem. Phys.} \textbf{\bibinfo{volume}{117}},
  \bibinfo{pages}{5859} (\bibinfo{year}{2002}).

\bibitem[{\citenamefont{Karasevskii and Holzapfel}(2003)}]{ka03}
\bibinfo{author}{\bibfnamefont{A.~I.} \bibnamefont{Karasevskii}}
  \bibnamefont{and} \bibinfo{author}{\bibfnamefont{W.~B.}
  \bibnamefont{Holzapfel}}, \bibinfo{journal}{Phys. Rev. B}
  \textbf{\bibinfo{volume}{67}}, \bibinfo{pages}{224301}
  (\bibinfo{year}{2003}).

\bibitem[{\citenamefont{Lawler et~al.}(2004)\citenamefont{Lawler, Chang, and
  Shirley}}]{la04}
\bibinfo{author}{\bibfnamefont{H.~M.} \bibnamefont{Lawler}},
  \bibinfo{author}{\bibfnamefont{E.~K.} \bibnamefont{Chang}}, \bibnamefont{and}
  \bibinfo{author}{\bibfnamefont{E.~L.} \bibnamefont{Shirley}},
  \bibinfo{journal}{Phys. Rev. B} \textbf{\bibinfo{volume}{69}},
  \bibinfo{pages}{174104} (\bibinfo{year}{2004}).

\bibitem[{\citenamefont{Pollock et~al.}(1972)\citenamefont{Pollock, Bruce,
  Chester, and Krumhansl}}]{po72}
\bibinfo{author}{\bibfnamefont{E.~L.} \bibnamefont{Pollock}},
  \bibinfo{author}{\bibfnamefont{T.~A.} \bibnamefont{Bruce}},
  \bibinfo{author}{\bibfnamefont{G.~V.} \bibnamefont{Chester}},
  \bibnamefont{and} \bibinfo{author}{\bibfnamefont{J.~A.}
  \bibnamefont{Krumhansl}}, \bibinfo{journal}{Phys. Rev. B}
  \textbf{\bibinfo{volume}{5}}, \bibinfo{pages}{4180} (\bibinfo{year}{1972}).

\bibitem[{\citenamefont{Etters and Danilowicz}(1974)}]{et74}
\bibinfo{author}{\bibfnamefont{R.}~\bibnamefont{Etters}} \bibnamefont{and}
  \bibinfo{author}{\bibfnamefont{R.~L.} \bibnamefont{Danilowicz}},
  \bibinfo{journal}{Phys. Rev. A} \textbf{\bibinfo{volume}{9}},
  \bibinfo{pages}{1698} (\bibinfo{year}{1974}).

\bibitem[{\citenamefont{Gillan}(1988)}]{gi88}
\bibinfo{author}{\bibfnamefont{M.~J.} \bibnamefont{Gillan}},
  \bibinfo{journal}{Phil. Mag. A} \textbf{\bibinfo{volume}{58}},
  \bibinfo{pages}{257} (\bibinfo{year}{1988}).

\bibitem[{\citenamefont{Ceperley}(1995)}]{ce95}
\bibinfo{author}{\bibfnamefont{D.~M.} \bibnamefont{Ceperley}},
  \bibinfo{journal}{Rev. Mod. Phys.} \textbf{\bibinfo{volume}{67}},
  \bibinfo{pages}{279} (\bibinfo{year}{1995}).

\bibitem[{\citenamefont{Noya et~al.}(1996)\citenamefont{Noya, Herrero, and
  Ram\'{\i}rez}}]{no96}
\bibinfo{author}{\bibfnamefont{J.~C.} \bibnamefont{Noya}},
  \bibinfo{author}{\bibfnamefont{C.~P.} \bibnamefont{Herrero}},
  \bibnamefont{and}
  \bibinfo{author}{\bibfnamefont{R.}~\bibnamefont{Ram\'{\i}rez}},
  \bibinfo{journal}{Phys. Rev. B} \textbf{\bibinfo{volume}{53}},
  \bibinfo{pages}{9869} (\bibinfo{year}{1996}).

\bibitem[{\citenamefont{Landau and Lifshitz}(1980)}]{la80}
\bibinfo{author}{\bibfnamefont{L.~D.} \bibnamefont{Landau}} \bibnamefont{and}
  \bibinfo{author}{\bibfnamefont{E.~M.} \bibnamefont{Lifshitz}},
  \emph{\bibinfo{title}{Statistical Physics}} (\bibinfo{publisher}{Pergamon},
  \bibinfo{address}{Oxford}, \bibinfo{year}{1980}), \bibinfo{edition}{3rd} ed.

\bibitem[{\citenamefont{Timms et~al.}(1996)\citenamefont{Timms, Evans,
  Boninsegni, Ceperley, Mayers, and Simmons}}]{ti96}
\bibinfo{author}{\bibfnamefont{D.~N.} \bibnamefont{Timms}},
  \bibinfo{author}{\bibfnamefont{A.~C.} \bibnamefont{Evans}},
  \bibinfo{author}{\bibfnamefont{M.}~\bibnamefont{Boninsegni}},
  \bibinfo{author}{\bibfnamefont{D.~M.} \bibnamefont{Ceperley}},
  \bibinfo{author}{\bibfnamefont{J.}~\bibnamefont{Mayers}}, \bibnamefont{and}
  \bibinfo{author}{\bibfnamefont{R.~O.} \bibnamefont{Simmons}},
  \bibinfo{journal}{J. Phys.: Condens. Matter} \textbf{\bibinfo{volume}{8}},
  \bibinfo{pages}{6665} (\bibinfo{year}{1996}).

\bibitem[{\citenamefont{Herrero}(2003{\natexlab{b}})}]{he03b}
\bibinfo{author}{\bibfnamefont{C.~P.} \bibnamefont{Herrero}},
  \bibinfo{journal}{Phys. Rev. B} \textbf{\bibinfo{volume}{68}},
  \bibinfo{pages}{172104} (\bibinfo{year}{2003}{\natexlab{b}}).

\bibitem[{\citenamefont{Flubacher et~al.}(1961)\citenamefont{Flubacher,
  Leadbetter, and Morrison}}]{fl61}
\bibinfo{author}{\bibfnamefont{P.}~\bibnamefont{Flubacher}},
  \bibinfo{author}{\bibfnamefont{A.~J.} \bibnamefont{Leadbetter}},
  \bibnamefont{and} \bibinfo{author}{\bibfnamefont{J.~A.}
  \bibnamefont{Morrison}}, \bibinfo{journal}{Proc. Phys. Soc. (London)}
  \textbf{\bibinfo{volume}{A78}}, \bibinfo{pages}{1449} (\bibinfo{year}{1961}).

\bibitem[{\citenamefont{Somoza and Fenichel}(1971)}]{so71}
\bibinfo{author}{\bibfnamefont{E.}~\bibnamefont{Somoza}} \bibnamefont{and}
  \bibinfo{author}{\bibfnamefont{H.}~\bibnamefont{Fenichel}},
  \bibinfo{journal}{Phys. Rev. B} \textbf{\bibinfo{volume}{3}},
  \bibinfo{pages}{3434} (\bibinfo{year}{1971}).

\bibitem[{\citenamefont{Hardy et~al.}(1998)\citenamefont{Hardy, Lacks, and
  Shukla}}]{ha98}
\bibinfo{author}{\bibfnamefont{R.~J.} \bibnamefont{Hardy}},
  \bibinfo{author}{\bibfnamefont{D.~J.} \bibnamefont{Lacks}}, \bibnamefont{and}
  \bibinfo{author}{\bibfnamefont{R.~C.} \bibnamefont{Shukla}},
  \bibinfo{journal}{Phys. Rev. B} \textbf{\bibinfo{volume}{57}},
  \bibinfo{pages}{833} (\bibinfo{year}{1998}).

\bibitem[{\citenamefont{Herrero and Ram\'{\i}rez}(2001)}]{he01}
\bibinfo{author}{\bibfnamefont{C.~P.} \bibnamefont{Herrero}} \bibnamefont{and}
  \bibinfo{author}{\bibfnamefont{R.}~\bibnamefont{Ram\'{\i}rez}},
  \bibinfo{journal}{Phys. Rev. B} \textbf{\bibinfo{volume}{63}},
  \bibinfo{pages}{024103} (\bibinfo{year}{2001}).

\bibitem[{\citenamefont{Lotrich and Szalewicz}(1997)}]{lo97}
\bibinfo{author}{\bibfnamefont{V.~F.} \bibnamefont{Lotrich}} \bibnamefont{and}
  \bibinfo{author}{\bibfnamefont{K.}~\bibnamefont{Szalewicz}},
  \bibinfo{journal}{Phys. Rev. Lett.} \textbf{\bibinfo{volume}{79}},
  \bibinfo{pages}{1301} (\bibinfo{year}{1997}).

\bibitem[{\citenamefont{Tang and Toennis}(1984)}]{ta84}
\bibinfo{author}{\bibfnamefont{K.~T.} \bibnamefont{Tang}} \bibnamefont{and}
  \bibinfo{author}{\bibfnamefont{J.~P.} \bibnamefont{Toennis}},
  \bibinfo{journal}{J. Chem. Phys.} \textbf{\bibinfo{volume}{80}},
  \bibinfo{pages}{3726} (\bibinfo{year}{1984}).

\bibitem[{\citenamefont{Anderson et~al.}(1973)\citenamefont{Anderson, Fugate,
  and Swenson}}]{an73}
\bibinfo{author}{\bibfnamefont{M.~S.} \bibnamefont{Anderson}},
  \bibinfo{author}{\bibfnamefont{R.~Q.} \bibnamefont{Fugate}},
  \bibnamefont{and} \bibinfo{author}{\bibfnamefont{C.~A.}
  \bibnamefont{Swenson}}, \bibinfo{journal}{J. Low Temp. Phys.}
  \textbf{\bibinfo{volume}{10}}, \bibinfo{pages}{345} (\bibinfo{year}{1973}).

\bibitem[{\citenamefont{Anderson and Swenson}(1975)}]{an75}
\bibinfo{author}{\bibfnamefont{M.~S.} \bibnamefont{Anderson}} \bibnamefont{and}
  \bibinfo{author}{\bibfnamefont{C.~A.} \bibnamefont{Swenson}},
  \bibinfo{journal}{J. Phys. Chem. Solids} \textbf{\bibinfo{volume}{36}},
  \bibinfo{pages}{145} (\bibinfo{year}{1975}).

\bibitem[{\citenamefont{Noya et~al.}(1997)\citenamefont{Noya, Herrero, and
  Ram\'{\i}rez}}]{no97}
\bibinfo{author}{\bibfnamefont{J.~C.} \bibnamefont{Noya}},
  \bibinfo{author}{\bibfnamefont{C.~P.} \bibnamefont{Herrero}},
  \bibnamefont{and}
  \bibinfo{author}{\bibfnamefont{R.}~\bibnamefont{Ram\'{\i}rez}},
  \bibinfo{journal}{Phys. Rev. B} \textbf{\bibinfo{volume}{56}},
  \bibinfo{pages}{237} (\bibinfo{year}{1997}).

\bibitem[{\citenamefont{Herrero}(2000)}]{he00a}
\bibinfo{author}{\bibfnamefont{C.~P.} \bibnamefont{Herrero}},
  \bibinfo{journal}{Phys. Status Solidi B} \textbf{\bibinfo{volume}{220}},
  \bibinfo{pages}{857} (\bibinfo{year}{2000}).

\end{thebibliography}

\end{document}